\documentclass[%
 aip,
 amsmath,amssymb,
 reprint,%
]{revtex4-1}

\usepackage{graphicx}
\usepackage{dcolumn}
\usepackage{bm}
\usepackage{placeins}

\usepackage[utf8]{inputenc}
\usepackage[T1]{fontenc}
\usepackage{mathptmx}
\usepackage{xcolor}
\usepackage{float}
\usepackage{soul}
\usepackage{chemformula}

\begin{document}

\newcommand{\mapbi}{$\mathrm{MAPbI_3}$}

\preprint{AIP/123-QED}

\title{Structure and Binding in Halide Perovskites: Analysis of Static and Dynamic Effects from Dispersion-Corrected Density Functional Theory}

\author{H. Beck}
\author{C. Gehrmann}%
\author{D. A. Egger}
 \email{david.egger@physik.uni-regensburg.de}
\affiliation{ 
Institute of Theoretical Physics, University of Regensburg, Universitätsstraße 31, 93053 Regensburg
}%

\date{\today}

\begin{abstract}
	We investigate the impact of various levels of approximation in density functional theory calculations for the structural and binding properties of the prototypical halide perovskite \mapbi. Specifically, we test how the inclusion of different correction schemes for including dispersive interactions, and how in addition using hybrid density functional theory, affects the results for pertinent structural observables by means of comparison to experimental data. {In particular, the impact of finite temperature on the lattice constants and bulk modulus, and the role of dispersive interactions in calculating them, is examined by using molecular dynamics based on density functional theory.} Our findings confirm previous theoretical work showing that including dispersive corrections is crucial for accurate calculation of structural and binding properties of \mapbi{}. They furthermore highlight that using a computationally much more expensive hybrid density functional has only minor consequences for these observables. This allows for suggesting the use of semilocal density functional theory, augmented by pairwise dispersive corrections, as a reasonable choice for structurally more complicated calculations of halide perovskites. Using this method, we perform molecular dynamics calculations and discuss the dynamic effect of molecular rotation on the structure of and binding in \mapbi, which allowed for rationalizing microscopically the simultaneous occurrence of cubic octahedral symmetry and MA disorder.
\end{abstract}

\maketitle

\section{Introduction}
	In recent years, the hope of cheap, solution-based solar cells with efficiencies above 20\% has fueled the scientific research in halide perovskites (HaPs).
    \cite{David1_1, David1_2, David1_3, David1_4, David1_5, David1_6, David1_7}    
    HaPs exhibit many interesting optoelectronic properties that make them ideal candidates for future semiconductor based applications, yet some of their structural and dynamic properties remain of central scientific interest since they also present major stumbling blocks in the development of stable devices. Fundamentally, HaPs can crystallize in the typical perovskite structure $ABX_3$, where the $B$-site is a heavy metallic element such as lead, and halide ions ($X$) form octahedra centered around the $B$-site. 
    \cite{David2}
    In hybrid organic-inorganic HaPs, the voids between the inorganic scaffolds are occupied by the organic $A$-site (in this work methylammonium, MA). As is typical for perovskites, HaPs undergo phase transitions with changes in temperature. The prototypical variant \mapbi{} has an orthorhombic symmetry up to 162~K, where it changes to a tetragonal structure. The second phase transition occurs at 327~K, where it switches to a cubic crystal structure.
    \cite{Poglitsch87, Stoumpos13}
    Importantly, HaPs are compared to other well established semiconductors
    \cite{David4}
    and perovskite materials very soft mechanically,
    \cite{Feng14, David5_2, Rakita15, David5_4, David5_5, David5_6}
    which is interesting fundamentally and could become problematic when they are exposed to the operating conditions of commercially used solar cells. Furthermore, experimental and theoretical studies have also shown that the structure and binding in HaPs can be modified already by applying relatively low pressure. 
    \cite{Capitani16, David6_2, David6_3}
Therefore, it is interesting to study the interrelation of binding and structural modifications in HaPs, which can be induced by a change in temperature or applied pressure, and may have important consequences for their optoelectronic properties.
    \newline
    Such insight can be generated from first-principles based computations. However, from a theoretical point of view there are a number of physical effects and properties, all of which could play a role regarding the structure of and binding in HaPs. First, hybrid HaPs consist of highly polarizable atoms such as iodine and lead and contain hydrogen atoms bound to electronegative species (e.g. nitrogen); hence, from simple chemical arguments dispersive interactions, such as van-der-Waals (vdW) and hydrogen bonding, can be expected to play an important role in HaPs. {Several computational studies have shown that dispersive interactions are crucial in static calculations of the lattice constants and mechanical properties of HaPs.} 
    \cite{Feng14, David7_2, Egger14, David7_4, David7_5,Faghihnasiri17, David5_5, Motta16, Egger18}
    However, as to how their treatment for HaPs should best be implemented in density functional theory (DFT), the most widely used method for HaPs, is still an open question.\cite{Kronik14, Hermann17} {In particular, a recent study has concluded that correcting for dispersive interactions in calculations of HaPs does not improve their description.}
    \cite{Bokdam17}
Second, Kohn-Sham DFT functionals, such as the frequently used generalized gradient approximation (GGA), cannot accurately describe the band gap of semiconductors even in principle,
    \cite{David8_1, David8_2}
    which can be improved by applying computationally more costly hybrid functionals in the generalized Kohn-Sham framework.
    \cite{David9,Zhang11}
For classical inorganic semiconductors, screened hybrid functionals were shown to improve also the description of lattice constants compared to GGA-based approaches.
	\cite{David10_1, David10_2, David10_3}
    The question that naturally arises then is whether a similar improvement is found when using a hybrid functional for calculating the structure of HaPs, {as has been suggested recently.}\cite{Bokdam17}
    Third, it is well established that spin-orbit coupling (SOC) due to the presence of the lead atom in \mapbi{} leads to strong modifications in the electronic structures, i.e., it lifts the degeneracy of the conduction and valence band and lowers the band gap.
    \cite{David11,Whalley17}
    Lastly, since HaPs are mechanically soft, they exhibit large structural dynamical effects, such as massive ionic displacements and molecular rotation around room temperature,
    \cite{David12_1, Egger16}
    which is the relevant scenario for device applications. Therefore, it is important to investigate the consequences of these unusual structural dynamical effects on the pertinent structural properties of HaPs. {To the best of our knowledge, the impact of finite temperature on the lattice constants and mechanical properties, and the role of dispersive interactions in calculating them, has not been addressed by means of DFT-based molecular dynamics (MD).}
    \newline
    In this study, we first investigate how the choice of vdW correction and DFT functional affects the results of calculations for structure and binding in the prototype \mapbi. To this end, we compare data obtained in static DFT calculations for a primitive cubic unit cell to experimental ones, using various computational approaches. The most promising choice of method is then used to explore how different MA orientations impact structural properties in static and {finite-temperature MD calculations} of \mapbi.

\section{Methods and Computational Setup}

	A satisfactory treatment of dispersion forces within conventional approximate Kohn-Sham DFT functionals requires dispersive correction schemes. In our calculations, we considered the Tkatchenko-Scheffler method with regular Hirshfeld partitioning (TS)
    \cite{Tkatchenko09}
    and with an iterative Hirshfeld partitioning (HI)
    \cite{Bucko14, Bucko13}
as well as the many-body dispersion (MBD) method.
	\cite{David15, Ambrosetti14, Bucko16}
The TS and HI methods include pairwise dispersive interactions between atoms in the crystal. The HI scheme expands the Hirshfeld partitioning with an iterative process that results in a more realistic allocation of charges for strongly ionic systems.
\cite{Bucko14}
To include the screening of dispersive interactions that occurs in a many-body system, we apply the MBD method based on the random phase expression of the correlation energy, as developed by Tkatchenko et al.
	\cite{David15}
\\    
    For the comparison of results obtained by using a GGA functional and a hybrid functional, we apply two very common variants in the context of solid-state calculations, namely the PBE
    \cite{Perdew96}
    and HSE functionals.
    \cite{Heyd03, David18_2}
    In the HSE functional, the short-range exchange energy is calculated as a mix of PBE and exact exchange,
    \cite{Heyd03, David18_2}
    which was often found to improve the description of electronic-structure and structural properties of semiconductors. 
    \cite{David20_1, David10_1, David10_2, David10_3}
    SOC describes the interaction of the spin angular momentum with the orbital momentum, and it plays a crucial role in systems with heavy elements such as lead. In our calculations, it was described {fully self-consistently} within the framework of non-collinear magnetism,
    \cite{David21}
    in conjunction with the various DFT methods applied here.    
 
    In order to obtain structural and mechanical parameters, we use an equation of state that accurately describes the macroscopic properties of the crystal. One of the most frequently used ones is the Birch-Murnaghan equation of state (BMEOS).
    \cite{Birch,Murnaghan,Fu83}
    It describes the free energy of a solid as a function of the volume of the unit cell, $E(V)$, at isothermal conditions as:
    \begin{eqnarray}
    	E(V) = E_0 + \frac{B_0V}{B^\prime_0} \left( \frac{(V_0/V)^{B^\prime_0}}{B^\prime_0 - 1} +1 \right) - \frac{V_0B_0}{B^\prime_0 -1}.\label{BMEOS}
        \end{eqnarray}
    $E_0$ is the free energy at the equilibrium volume, $V_0$, and $B_0$ and $B^\prime_0$ are the bulk modulus and its derivative, respectively. In order to obtain these parameters, we first calculate the free energy of the static system at eight equidistant volumes between 90\% and 111\% of the experimental value
    \cite{Stoumpos13}
    and fit Eq.~\ref{BMEOS} using these eight values (tests with more data points did not show any significant improvement) with the method of least squares. Note that while the $E_0$ and  $B^\prime_0$ are required to fit Eq.~\ref{BMEOS}, they are not relevant for our discussion.
  
  \begin{figure}
	\includegraphics[width=\linewidth]{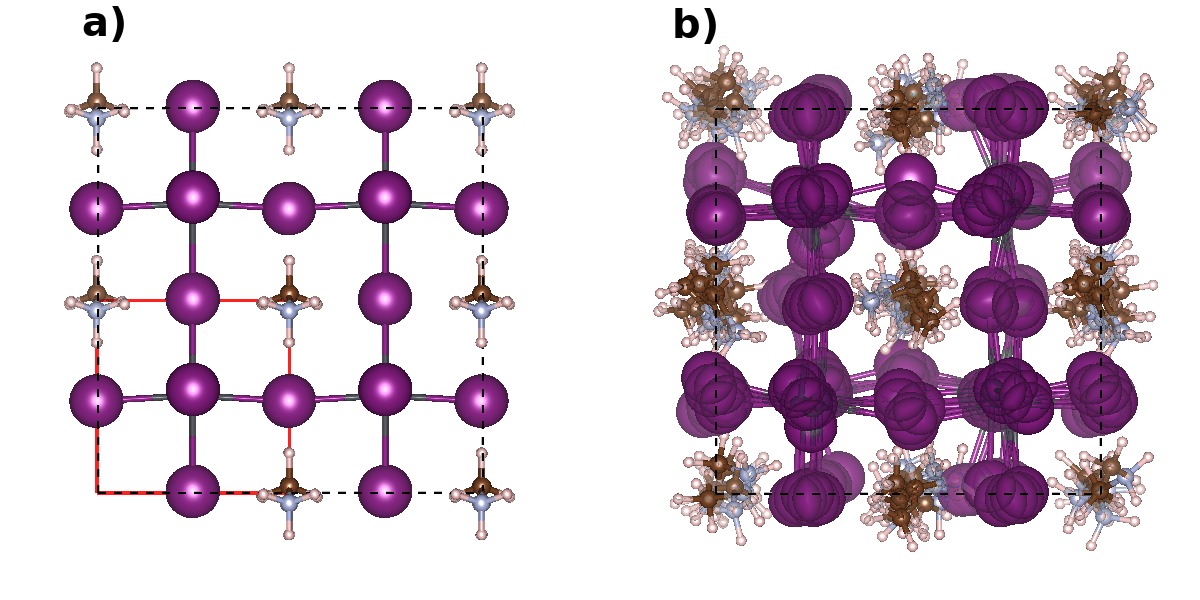}
    \caption{\label{combined} Schematic structural representations of the configurations of \mapbi{} that were used in the calculations: a) Primitive unit cell (red solid line) and a 2x2x2 supercell (black dashed line) used in static calculations; note that we have considered various orientations of MA, see text for details. b) Visualization of the structural changes in molecular dynamics (MD) calculation of a 2x2x2 supercell, which was obtained as an overlay of five structures separated by 200~fs along the 20~ps DFT-MD trajectory. Shown are carbon (brown), nitrogen (light blue), hydrogen (white), iodine (violet), and lead (gray) atoms. Atoms belonging to more than the computational cell are displayed for visual clarity.}
\end{figure}
  
    In our calculations, we first considered the cubic primitive unit cell of \mapbi{} as has been reported experimentally, see Fig.~\ref{combined}a.
    \cite{Stoumpos13}
    To be able to also investigate the effect of MA orientation, we used a supercell that consists of eight unit-cells stacked together in a 2x2x2 pattern (see Fig.~\ref{combined}a), with the MA ions orientated in parallel or anti-parallel to each other. {In this way, we can test the effect of the assumption that MA molecules are oriented perfectly in parallel throughout the material on computing the structural and binding properties of \mapbi{}.} Note that in these calculations, the angle between the MA molecules was fixed during the atomic relaxation. To fit Eq.~\ref{BMEOS} with the data obtained in the MD calculations, the average free energy along a trajectory of 20~ps was calculated for each volume point, which was used, together with the standard deviation as the uncertainty, in the fit of Eq. \ref{BMEOS}. We note that since the fit errors of $B^\prime$ were quite large in the case of the MD calculations, in fitting Eq.~\ref{BMEOS} we set $B^\prime$ to the value obtained in the static calculation of the unit cell.

	\begin{table*}
    	\caption{\label{static_unitcell} Volume of the primitive unit cell, $V_0$, and bulk modulus, $B_0$, obtained by fitting the DFT-calculated data with Eq.~\ref{BMEOS}, using various methods applied to the primitive unit-cell of \mapbi.}
    	\begin{ruledtabular}
    		\begin{tabular}{lccccccccccc}
				& PBE & PBE+TS & PBE+MBD & PBE+HI & HSE & HSE+TS & HSE+MBD & HSE+HI & PBE+SOC & PBE+TS+SOC & Experiment \\
                ine
                $V_0$ [\AA$^3$] \footnote{Errors are between 0.1 and 0.4~\AA$^3$} & 272.9 & 256.2 & 257.1 & 262.0 & 266.6 & 252.9 & 252.1 & 258.1 & 274.4 & 256.3 & 247-253 \cite{Stoumpos13,Baikie13,Feng14}\\
                $B_0$ [GPa] \footnote{Errors are between 0.2 and 0.4~GPa}& 10.6 & 15.7 & 13.6 & 14.3 & 11.1 & 16.4 & 14.7 & 14.2 & 9.6 & 15.0 & 12-16 \cite{Ferreira18,Rakita15}\\
           	\end{tabular}
    	\end{ruledtabular}
    \end{table*}
    
    The DFT calculations were performed with a plane-wave basis (cutoff energy: 400 eV) and the projector-augmented wave (PAW) method,
    \cite{David24}
    as implemented in the VASP code.
    \cite{David25}
    For unit cell calculations, a 6x6x6 grid of k-points centered around the $\Gamma$-point was used, for static and dynamic calculations that considered the supercell, the grid was reduced to a sampling of 3x3x3. For each static calculation, the ionic coordinates were relaxed, using the PBE functional and the respective dispersive correction scheme, with the Gadget tool in internal coordinates,
    \cite{David26}
    before calculating the energies to fit Eq.~\ref{BMEOS}. In the calculations applying the HSE functional or SOC, the respective PBE-based geometry was used to reduce computational costs. For all static calculations, the geometry was considered relaxed when the forces acting  one the atoms were below 10~meV per \AA. In the canonical (NVT) MD simulations, a \mapbi{} 2x2x2 supercell with randomly orientated MA was used as a starting point for the structural geometry, {in line with the recommendations provided by Lahnsteiner et al.} \cite{Lahnsteiner16}
    It was then equilibrated for 5~ps and computed for 20~ps in time steps of 1~fs at a temperature of 400~K. {This protocol is sufficient to compute the impact of the dynamic nuclear effects on the structure and binding despite the limited supercell size} \cite{Lahnsteiner16}{ and trajectory length. The latter was checked explicitly by adding another 10~ps to the simulation, which resulted in only insignificant changes of the calculated observables.} Schematic representations of the structures were generated with VESTA. \cite{vesta}

\section{Results}

\begin{figure}
    	\includegraphics[width=1.0\linewidth]{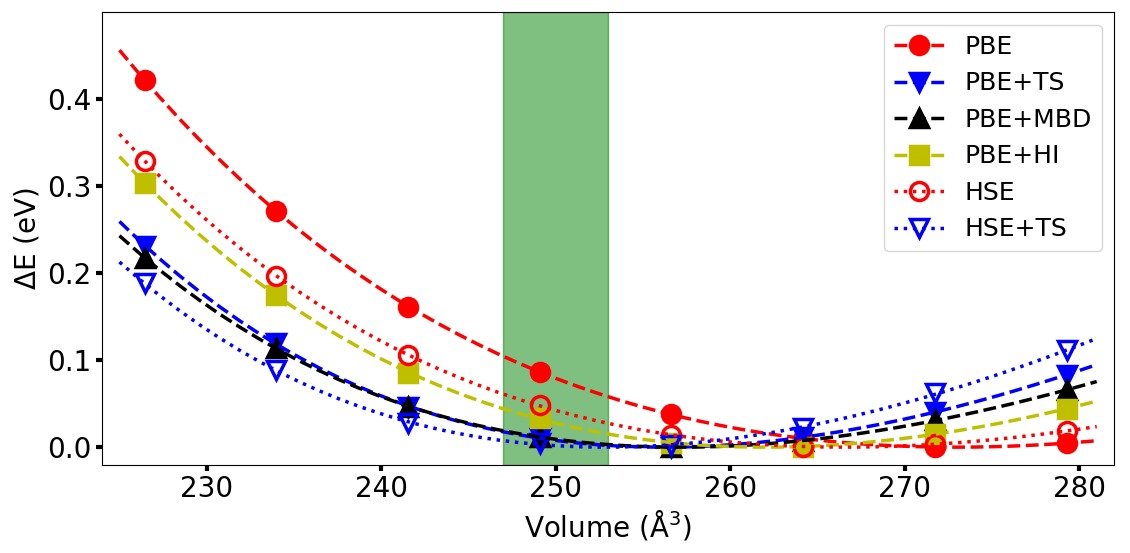}
        \caption{\label{vdw_comp} Energy change as a function of unit-cell volume, $\Delta E(V) = E(V)- E_0$, for the static calculations of the primitive \mapbi{} unit cell, using the PBE and HSE functionals augmented by different dispersion correction schemes. Also shown are fits to Eq.~\ref{BMEOS} and the range of experimental results for the volumes (green-shaded area). It is noted that a larger slope in the fit corresponds to a larger value of $B_0$.}
    \end{figure}

Fig.~\ref{vdw_comp} shows the energy change due to volume change, i.e., $\Delta E(V) = E(V)- E_0$, for the static calculations of the \mapbi{} unit cell, calculated by using the PBE and HSE functionals augmented by different dispersion corrections. The optimized volume ($V_0$) and bulk modulus ($B_0$), obtained from the fit, are listed in Table~\ref{static_unitcell}. Considering the PBE data, it can readily be seen that using dispersive corrections has a large impact on the result. Furthermore, it was found that the results from the TS and MBD methods are very similar, and provide a $V_0$ of 256.2~\AA$^3$ (PBE+TS) and 257.1~\AA $^3$ (PBE+MBD) which are close to the experimental range of 247-253~\AA$^3$.
In contrast, the bare PBE calculations are far off (272.9~\AA$^3$) the experimental range, in agreement with literature findings,
 \cite{Feng14, David7_2, Egger14, David7_4, David7_5,Faghihnasiri17}
and the PBE+HI (262.0~\AA$^3$) results basically lie in between the results of bare PBE and PBE+TS/PBE+MBD. The calculated $B_0$ data show similar trends, i.e., the experimentally reported values (12-16~GPa)
agree well with the PBE+TS, PBE+MBD, and PBE+HI results, but deviate more from the bare PBE value. From these data, one can see that the TS dispersive correction scheme with a regular Hirshfeld partitioning and the MBD approach perform best in reproducing structural and binding properties measured in experiments. {It should be noted that the experimental data were recorded at elevated temperatures, at least at $T\approx$330~K, while the calculations were performed at 0~K and do not address thermal expansion, an effect we discuss below based on our results obtained from finite-temperature MD calculations.}

    Fig.~\ref{vdw_comp} also shows the results for regular and TS-corrected calculations using the HSE functional (see Table~\ref{static_unitcell} for full dataset using the other correction schemes). While there is some visible difference between the PBE and HSE calculations without dispersive corrections, the bare HSE value for $V_0$ is still quite off the experimental result, in stark contrast to findings reported for conventional inorganic semiconductors.
\cite{David10_1, David10_2, David10_3}    
    Considering the dispersion-corrected HSE data, it is found that these compare almost equally well to experiment as their PBE counterparts. Importantly, the optimized $V_0$ value using the PBE+TS, PBE+MBD, HSE+TS, and HSE+MBD are all within $< 5$~\AA$^3$, which is smaller than the range of experimentally reported values. Hence, we find that the improvement due to using the HSE functional is minor compared to using dispersive corrections, which is thus found to be the essential computational ingredient for obtaining accurate structural and binding properties of \mapbi. From these findings, we argue that using PBE+TS provides a reasonable choice for calculating structural and binding properties of HaPs such as \mapbi. For completeness, we note that calculations including SOC showed very little difference to those without, as can be seen in Table~\ref{static_unitcell}.

    It is well-known that the electronic interaction between MA and the inorganic ions in \mapbi{} is minor, since the electronic states of MA are energetically not close to the band edges. Nevertheless, electrostatic and dispersive interactions
between the inorganic ions and MA could still be important for the structure and binding of \mapbi. In order to test this, we varied the MA orientation in a 2x2x2 supercell, considering the extreme cases of either perfectly parallel or antiparallel MA molecules. {It is noted that the former scenario is equivalent to considering a primitive cubic unit cell of \mapbi{} containing one MA unit.} When calculating the structural parameters of the 2x2x2 supercell with different MA orientations, we limit the calculations to the bare PBE and the PBE+TS approach, since neither using HSE nor including SOC has shown a significant improvement that would justify the increase in computational cost {(see above and Table}~\ref{static_unitcell}). Furthermore, using the TS and MBD dispersive correction scheme provided essentially equal results, but the computational costs associated with the MBD method increase more rapidly when the number of atoms becomes larger compared to the TS scheme.
   
\begin{figure}
    	\includegraphics[width=.49\linewidth]{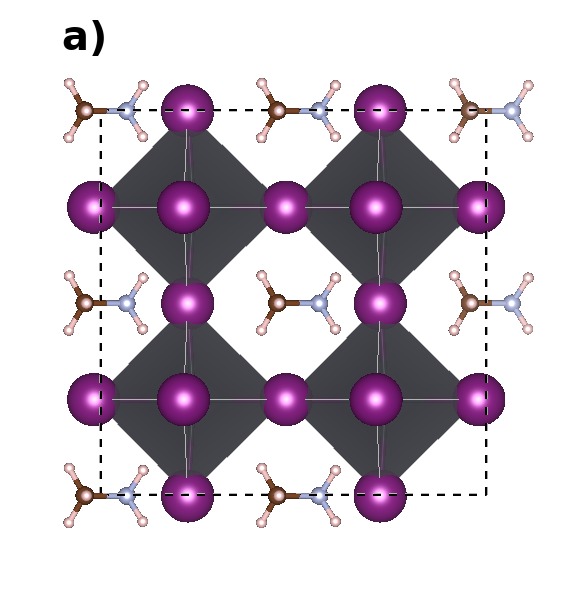}
        \includegraphics[width=.49\linewidth]{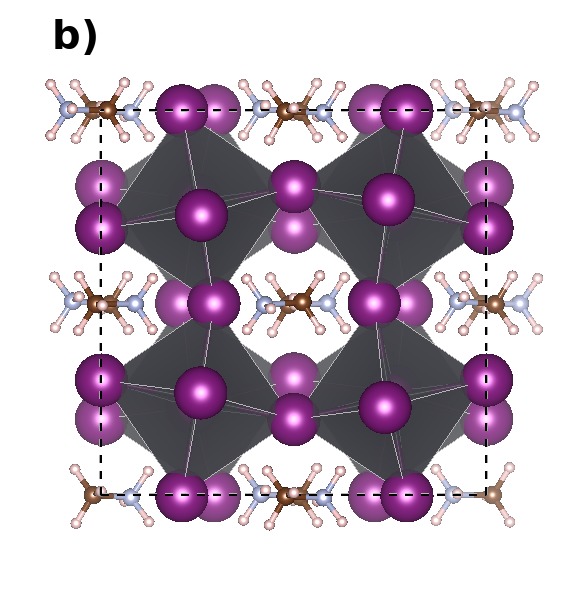}
        \caption{\label{supercell_MA}Schematic representations of the supercells with optimized atomic geometries obtained with PBE+TS at a volume of $256.6$~\AA$^3$, for the case of parallel (a) and antiparallel MA orientation (b). The x-z plane is shown, and atoms belonging to more than the computational cell are displayed for visual clarity}
    \end{figure}

   The first relevant observation in the data shown in Table~\ref{static_supercell} is that the results of the supercell calculations considering the parallel MA orientation are, within the fitting error, identical to the results obtained with the calculations for the primitive unit cell, as expected. However, a noticeable difference occurs in both the structure (see Fig.~\ref{supercell_MA}) and the optimized parameters (see Table~\ref{static_supercell}) when the MA orientation is changed to the other extreme of perfectly antiparallel MA molecules. In the case of the parallel MA orientation, at all volumes the octahdra retain cubic symmetry, whereas in the antiparallel orientation they tilt strongly (see Fig.~\ref{supercell_MA}). Indeed, this effect is important, since for the fit of the supercell data calculated with bare PBE we find that $V_0/8$ changes from 272.7\AA$^3$, for the parallel orientation, to 265.8\AA$^3$, for the antiparallel one. The same trend is confirmed in the PBE+TS calculations, although the differences are smaller. Hence, the interaction between the inorganic cage and the MA molecules is important for the structure and binding in \mapbi. Since our DFT calculations show that the antiparallel orientation is preferred in ~\mapbi, i.e., the free energy is consistently lower for the antiparallel MA orientation at all eight volumes, this requires further investigations.
    
    \begin{table}
    	\caption{\label{static_supercell}Cell volume, $V_0$, and bulk modulus, $B_0$, obtained by fitting the DFT-calculated data with Eq.~\ref{BMEOS}, using PBE and PBE+TS applied to a 2x2x2 supercell of \mapbi{} with differently orientated MA molecules. Note that in order to improve the comparison, we here report $V_0/8$.}
        \begin{ruledtabular}
            \begin{tabular}{lcc}
            	& parallel MA & anti-parallel MA \\
                ine                
                & \multicolumn{2}{c}{\textbf{PBE}} \\
                $V_0$ [\AA$^3$] \footnotemark[1]& 272.9 & 267.5  \\ 
                $B_0$ [GPa] \footnotemark[2]& 10.8 & 11.2 \\
                & \multicolumn{2}{c}{\textbf{PBE+TS}} \\
                $V_0$ [\AA$^3$] \footnotemark[1] & 256.3 & 253.8 \\ 
                $B_0$ [GPa] \footnotemark[2]& 15.7 & 14.7 \\
            \end{tabular}
            \footnotetext[1]{Errors are between 0.1 and 0.3~\AA$^3$}
            \footnotetext[2]{Errors are between 0.4 and 0.4~GPa}
        \end{ruledtabular}
    \end{table}

    \begin{table}
    	\caption{\label{MD_results}
Cell volume, $V_0$, and bulk modulus, $B_0$, obtained by fitting the DFT-calculated data with Eq.~\ref{BMEOS}, using PBE and PBE+TS applied to a 2x2x2 supercell of \mapbi{} computed along an MD trajectory of 20~ps. Note that in order to improve the comparison, we here report $V_0/8$.}
       	\begin{ruledtabular}
          	\begin{tabular}{lcc}
            	& PBE & PBE+TS \\
                ine
            	$V_0$ [\AA$^3$] \footnotemark[1] & 267.3 & 248.1 \\ 
            	$B_0$ [GPa] \footnotemark[1] & 8.8 & 13.0 \\
            \end{tabular}
            \footnotetext[1]{Errors are 0.1~\AA$^3$}
            \footnotetext[2]{Errors are between 0.6 and 0.7~GPa}
        \end{ruledtabular}
	\end{table}

    \begin{figure}
    	\includegraphics[width=\linewidth]{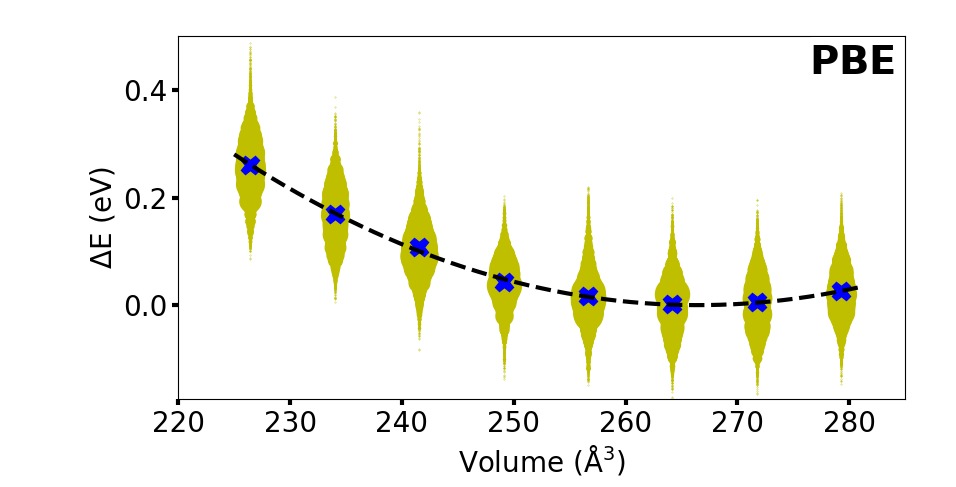}
        \includegraphics[width=\linewidth]{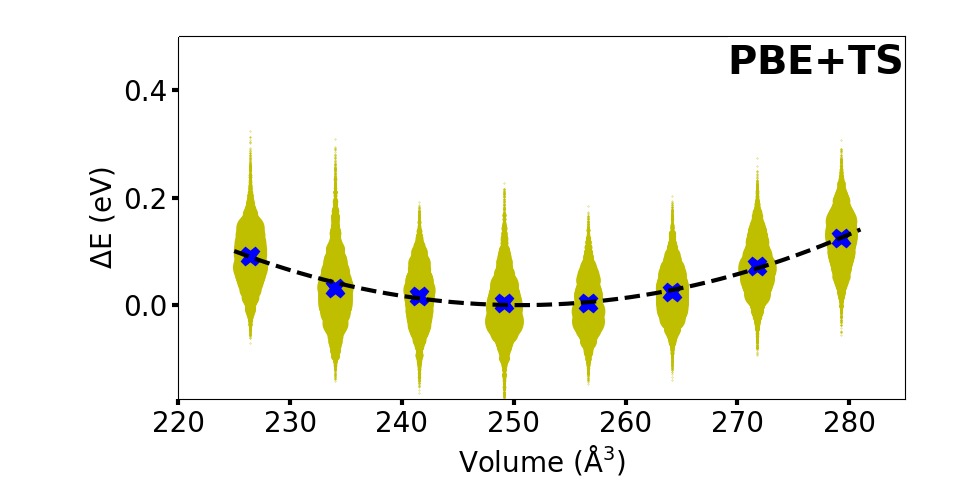}
        \caption{\label{md_bmeos}
        Energy change as a function of unit-cell volume, $\Delta E(V) = E(V)- E_0$, calculated by using MD-DFT with PBE (top) and PBE+TS (bottom) along a trajectory of 20~ps, shown as yellow dots, where the symbol size is given by number of occurrences in the MD run. The blue cross denotes the average $\Delta E(V)$, and the black dashed line is the fit according to Eq.~\ref{BMEOS}. Note that in order to improve the comparison, we here report $\Delta E/8$ as a function of $V/8$.}
    \end{figure}

    Hence, motivated by the finding that MA orientation impacts structure and binding in \mapbi, we performed fully unconstrained NVT MD calculations at 400~K. {This is relevant, since the MA unit was found to be only weakly bound in the cubic phase,} \cite{Chen15, Jingrui18} 
    {and hence undergoes rotational and translational motion at elevated temperatures, which could impact the energetics in the material dynamically. Furthermore,} the MD calculations allow for testing whether the effect of dispersive corrections seen for the static structures is still visible at elevated temperatures, {a comparison that has not been attempted previously but is} important for investigating dynamical effects in the structural interaction of \mapbi. Fig.~\ref{md_bmeos} shows the entire distribution as well as the mean value of the PBE- and PBE+TS-calculated change in free energy of \mapbi, $\Delta E$, determined for a 20~ps MD simulation, again fitted by Eq.~\ref{BMEOS}. The most apparent finding from Fig.~\ref{md_bmeos} is yet again the sizable difference between the PBE and PBE+TS curves, as also quantified by the $V_0$ parameters provided in Table~\ref{MD_results}. Furthermore, a slight decrease of $V_0$ is found in the MD calculations at 400~K compared to the static 0~K calculation, contrary to the expected thermal expansion. In regard to $B_0$, while in the case of PBE it is similar to the 0~K calculation of the primitive unit cell, for PBE+TS $B_0$ it is slightly lower than what was obtained in the static calculations. 
    
\begin{figure}
    	\includegraphics[width=.49\linewidth]{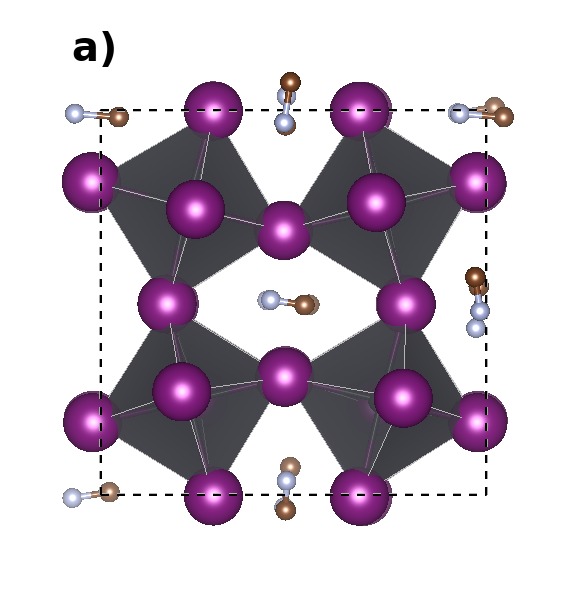}
        \includegraphics[width=.49\linewidth]{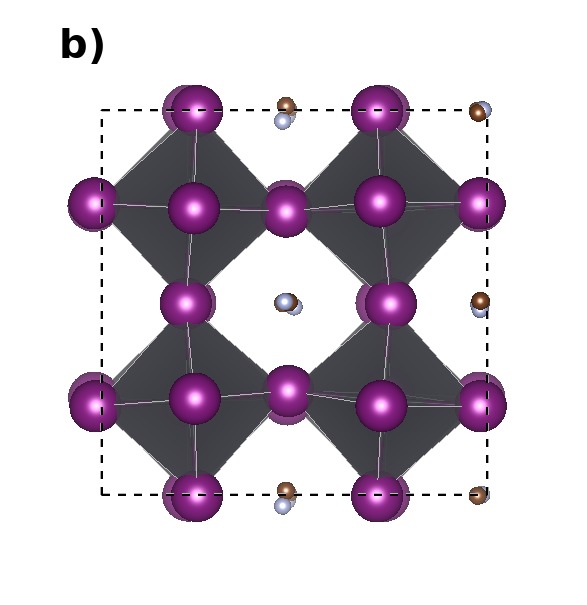}
        \caption{\label{md_poscars}Schematic representation of the time-averaged atomic positions along the MD trajectory, calculated with PBE+TS, for $V_0=231.0$\AA$^3$ (a) and $256.6$\AA$^3$ (b). Note that hydrogen atoms were omitted for clarity.}
    \end{figure}

    In order to better understand the origin of these findings, the average atomic positions along the {PBE+TS} MD trajectories were calculated. Fig.~\ref{md_poscars} shows that depending on the volume of the supercell, two different types of structures emerged. For the three smallest volumes considered in the MD, the octahedra were found to be tilted and the C-N bond of MA is still clearly visible in the average structure: successive MA molecules are oriented in parallel to each other along one direction and orthogonal to each other in the other two directions, aligning
    with the long axis of the rhombus created by the tilted octahedra. For the five larger considered volumes, on the other hand, the octahedra form an on average near perfect cubic symmetry, and the average carbon and nitrogen atoms are almost conjoined. The latter could either mean that there is absolutely no preferred direction of the MA molecules, i.e., it {rotates such that it is entirely disordered over the course of the 20~ps trajectory}, or that there are preferred directions which are equally occupied thermally.
    
\FloatBarrier    
\section{Discussion}

The results from the static unit cell calculations confirmed previous findings that showed the importance of including dispersive interaction in DFT calculations {for the structure and binding in \mapbi{}}.\cite{Feng14, David7_2, Egger14, David7_4, David7_5,Faghihnasiri17} {Here, we went beyond in testing a range of dispersive-correction schemes}: PBE calculations corrected by the TS scheme with the regular Hirshfeld partitioning and the MBD scheme showed good agreement with experimental data. The TS method with iterative Hirshfeld partitioning performed slightly worse than the regular TS and MBD schemes. {While the iterative Hirshfeld partitioning has indeed been shown to improve the description of ionic materials,}\cite{Bucko14, Bucko13} {\mapbi{} is a more complex case, since it is a hybrid organic-inorganic system that contains covalent bonds (in the MA molecule) as well as partially covalent-ionic bonds (in the inorganic framework)}. One perhaps surprising result is that the MBD method did not improve the results significantly compared to the TS method, even though one would expect that the iodine atoms, which account for most of the vdW-interactions,\cite{Egger14} are screened by the surrounding dielectric environment. {We considered a comparison of the $C_6$ parameters calculated from PBE+TS and PBE+MBD, which determine the dispersive correction in either case,}\cite{Kronik14,Hermann17}{ to further understand this result. We find that the changes are minor, on the order of $1~\%$ or smaller, and furthermore do not depend on the volume of the unit cell.} This implies that the screening is barely affected by the changes in distance in the calculations at different volumes, and thereby only leads to a constant energy shift. {Furthermore, we considered the higher-order contributions to the dispersive energy calculated with PBE+MBD, to find that indeed the second-order term is dominating. This could imply that the dispersive interactions in \mapbi{} are such that higher-order interactions are indeed negligible compared to the pairwise contribution, and also that the polarizability is largely isotropic (see ref. }\cite{Hermann17} 
{for further discussion).}
    
Furthermore, we found that using the hybrid functional HSE, which improves the description of the electronic structure, did not result in improvements for the optimized unit-cell volume and bulk modulus, once it was combined with dispersive corrections. Indeed, the results from the PBE+TS, PBE+MBD, HSE+TS, and HSE+MBD calculations are all within experimentally reported range for these two important structural parameters. Since {our data, presented in Table~}\ref{static_unitcell}{, also confirm the effect of SOC for the structure and binding to be minor, and since the PBE+TS approach was shown to be very accurate for structural and mechanical properties of multiple HaP crystals,} \cite{Egger14, David7_5, David5_5,David7_4, Motta16, Egger18} 
we conclude that using PBE+TS for investigating the structure and binding in static and dynamical calculations of HaPs is a reasonable choice.
   
{In this context, it is worth noting that} \citealt{Bokdam17} suggested the choice of DFT functional to have a bigger impact on the structural properties of \mapbi{} than the addition of dispersive corrections. While this is certainly true for electronic properties, and while our calculations showed some minor differences between the PBE and HSE results, {we have shown that} this clearly cannot diminish the important role of dispersive interactions in \mapbi. {We further note that the different conclusions of our work and} \citealt{Bokdam17} {could be related to the different points of reference used in either case: while we have chosen to consider experimental data on the lattice constants and bulk modulus,} \citealt{Bokdam17} {considered energies calculated in the random-phase approximation (RPA) as a reference. Indeed, RPA energies can be very accurate for solids and are broadly applicable, but it is worth noting that ``standard'' RPA suffers from known deficiencies, such as potentially incorrect descriptions of short-range correlation.} \cite{RPA_range,Hermann17}
    
    \begin{figure}
    	\includegraphics[width=\linewidth]{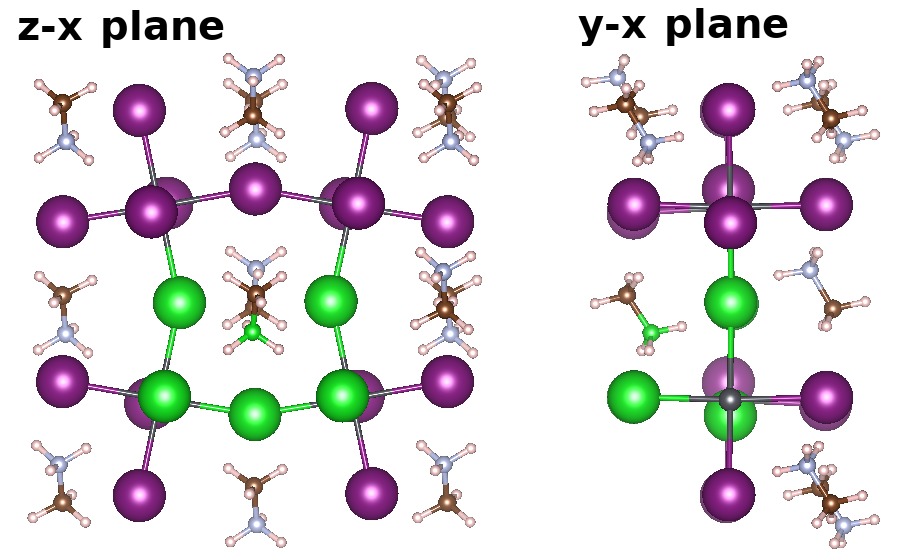}
        \caption{\label{antiparallel_mixed} Schematic representation of the interaction between the ammonium-end of MA and the surrounding iodine atoms; the relevant atoms are highlighted in green, and the z-x plane (left) and y-x plane (right) are shown. The structure is taken from the PBE+TS calculation of $V= 256.6$~\AA$^3$ as visualized in Fig.~\ref{supercell_MA}.}
    \end{figure}
    
Using the PBE+TS approach allowed for studying more complicated static and dynamic structural phenomena: First, we investigated the impact of MA orientation on the structural parameters of \mapbi, studying the extreme cases of either perfectly parallel or antiparallel MA molecules contained in a supercell. We found that only for the parallel orientation of MA do the octahedra retain cubic symmetry, while for antiparallel MA molecules they strongly tilt. Note that the lattice vectors were still constrained to the primitive unit cell, i.e., the volume change corresponds to a hydrostatic pressure in the system. Since the relative energy gain/cost due to these structural distortions depends on the unit-cell volume, this effect modified the obtained volume and bulk modulus. Hence, the interaction between MA and the inorganic atoms is important, especially because the calculations showed that the antiparallel orientations are actually energetically preferred.

The origin of these distortions is related to the interactions between the MA molecules and iodine atoms, since the partially positive ammonium-part of the MA interacts stronger with the partially negatively iodide ion than the methyl-side. In Fig.~\ref{antiparallel_mixed} we illustrate that the nitrogen atoms of MA interact mainly with five of the neighboring iodines, which results in a distortion of the octahedra such that the distances between the nitrogen and iodine atoms are maintained between 3.64 and 3.81~\AA. These interactions seem to be the main driving force behind the MA-induced distortions of the octahedra in the antiparallel case, which are energetically favorable for the system. Due to symmetry, these interactions are cancelled in the case of parallel MA orientation, and our geometry relaxations did not automatically adapt to the more favorable antiparallel szenario. This finding is also quite interesting in view of the fact that the experimentally-determined lattice symmetry of ~\mapbi{} above 327~K is almost perfectly cubic,\cite{Poglitsch87,Stoumpos13} despite the fact that different MA orientations can induce lower energy structures.

The implications of these findings can be fully understood by means of fully-unconstrained MD calculations at 400~K, since in these the MA molecules, as well as all other atoms, are allowed to move freely. The first important finding from these data was that inclusion of dispersive corrections is equally important to obtain reasonable structural parameters in MD calculations. Second, the results obtained from the MD simulations are quite surprising at first sight, since the unit-cell volume was found to be smaller than the one obtained from the 0~K static calculation. To understand this finding, consider that the time-scales associated with MA motion are much faster than the ones corresponding to the octahedral distortions of the heavy Pb-I cage, which is included in the MD but absent in the static calculations. Our analysis further showed that the average nuclear positions of MA are such that the carbon and nitrogen atoms are essentially conjoined at this temperature, meaning that MA is disordered, which is well known.\cite{Poglitsch87} Hence, the inorganic atoms essentialy respond to a time-averaged volume corresponding to the moving MA molecule. Due to the MA disorder, this volume is effectively smaller at 400~K than for a fixed pattern of MA orientations, allowing for shorter I-Pb-I bonds and hence smaller crystal volumes. Last but not least, this rationale also explains the finding from the MD data showing that the time-averaged octahedral symmetry is almost perfectly cubic at 400~K in agreement with experiment: MA disorder implies that all possible octahedral distortions induced by MA-iodine interactions are statistically equally likely. Therefore, the cubic perovskite structure is maintained as the most energetically favorable average crystal structure, exhibiting all the electronic and optical properties that render \mapbi{} so favorable for device applications.

Finally, the data contained in Fig.~\ref{md_poscars} showed that at smaller volumes the MD-averaged nuclear positions exhibit octahedral distortions together with a more preferred orientational order of MA. This makes sense, considering that smaller unit-cell volumes correspond to smaller voids between the octahedra, which increases the organic-inorganic interactions,\cite{David7_5} and partially hinders free MA motion.
Such tilting of the octahedra into an orthorhombic-like structure was also observed experimentally in pressure experiments\cite{Capitani16} for \mapbi, and the variation in the preferred direction for MA at different volumes was discussed also in a recent study reporting MD simulations.\cite{Lahnsteiner18} Therefore, in agreement with previous findings our study shows that even relatively mild changes in external pressure can result in large changes of the structure and binding in \mapbi.

\section{Conclusion}

 In summary, we investigated the impact of various levels of DFT-related approximations for calculations of the structural and binding properties of the prototypical HaP material \mapbi. Our tests considered the effects of including different dispersive correction schemes, applying a hybrid functional, including SOC, and also addressed the role of dynamic effects in MD calculations. The data confirmed previous theoretical work showing that dispersive corrections are important for accurate calculations of \mapbi, and also highlight that applying a computationally much more expensive hybrid functional improves the description of structural and mechanical properties by only a small amount. From this, we conclude that the use of a semilocal functional, augmented by pairwise dispersive interactions, is a suitable choice when computing more complicated static as well as structural dynamical phenomena in HaPs. Applying this methodology to DFT-based MD calculations of \mapbi, we analyzed the dynamic effect of molecular motion and its interplay with the structure of and binding in \mapbi. From this analysis, we could rationalize microscopically the simultaneous occurrence of a preferred cubic octahedral symmetry and MA disorder.
	
\section{Acknowledgements}

Funding provided by the Alexander von Humboldt Foundation in the framework of the Sofja Kovalevskaja Award endowed by the German Federal Ministry of Education and Research is acknowledged. The authors gratefully acknowledge the Gauss Centre for Supercomputing e.V. (www.gauss-centre.eu) for funding this project by providing computing time through the John von Neumann Institute for Computing (NIC) on the GCS Supercomputer JUWELS at Jülich Supercomputing Centre (JSC).

\section{References}

\bibliography{aipsamp}

\end{document}